\documentclass[aps,pre,twocolumn,nofootinbib,showpacs,longbibliography]{revtex4-1}
\usepackage{bm,amsmath,amssymb,graphicx}
\usepackage[colorlinks,linkcolor=blue,urlcolor=blue]{hyperref}
\begin{document}

\title{Natural Hamiltonian formulation of composite higher derivative theories}

\author{Hans Christian \"Ottinger}
\email[]{hco@mat.ethz.ch}
\homepage[]{http://www.polyphys.mat.ethz.ch/}
\affiliation{ETH Z\"urich, Department of Materials, Polymer Physics, HCP F 47.2,
CH-8093 Z\"urich, Switzerland}

\date{\today}

\begin{abstract}
If a higher derivative theory arises from a transformation of variables that involves time derivatives, a tailor-made Hamiltonian formulation is shown to exist. The details and advantages of this elegant Hamiltonian formulation, which differs from the usual Ostrogradsky approach to higher derivative theories, are elaborated for mechanical systems and illustrated for simple examples. Both a canonical space and a set of constraints emerge naturally from the transformation rule for the variables. In other words, the setting for quantization and the procedure for eliminating instabilities arise naturally.
\end{abstract}



\pacs{11.10.Ef}

\maketitle

\section{Introduction}
Most theories in physics are based on second-order ordinary or partial differential equations. However, there are good reasons for considering higher derivative theories, too. For example, in the context of gravity, higher derivative theories have been studied as alternative theories in their own right or as the expected result of renormalization, which is a much more general topic (see \cite{Stelle77,Stelle78,Krasnikov87,Beckeretal17} and references therein). Applications in particle physics include investigations of possible deviations of electroweak vector-boson self-interactions from the standard model (see \cite{GrosseKnetter94} and references therein).

Higher derivative theories are most easily introduced in a Lagrangian setting. If a Lagrangian $\bar{L}(\bar{q},\dot{\bar{q}},\ddot{\bar{q}})$ depends not only on the time derivatives $\dot{\bar{q}}$ of some configurational variables $\bar{q}$, but also on the second time derivatives $\ddot{\bar{q}}$, stationarity of the action $I = \int L dt$ typically leads to fourth order differential equations for $\bar{q}$ (except for linear dependence of $L$ on $\ddot{\bar{q}}$). The usual passage from the Lagrangian to the Hamiltonian framework clearly fails in the presence of second time derivatives in the Lagrangian. A generalization of the Hamiltonian approach to higher derivative theories was developed in a classical paper by Ostrogradsky \cite{Ostrogradsky1850} (a modern review with many illuminating comments can be found in an educational article by Woodard \cite{Woodard15}).

As the Ostrogradsky Hamiltonian is linear in some of the momenta and hence unbounded from below, instabilities are a major concern in higher derivative theories. Such instabilities can be suppressed by constraints \cite{Chenetal13} (or complexification \cite{RaidalVeermae17}, where the physical relevance of complex theories is not yet clear).

Recently we have introduced a class of higher derivative theories by considering a theory given by a standard Lagrangian $L(q,\dot{q})$  and a transformation $q=q(\bar{q},\dot{\bar{q}})$ \cite{hco235}. We say that $L(q,\dot{q})$ defines a workhorse theory and $q(\bar{q},\dot{\bar{q}})$ is a composition rule. A composite higher derivative theory is then introduced by the Lagrangian
\begin{equation}\label{Lcomposite}
   \bar{L}(\bar{q},\dot{\bar{q}},\ddot{\bar{q}}) = L\left( q(\bar{q},\dot{\bar{q}}),
   \frac{\partial q(\bar{q},\dot{\bar{q}})}{\partial \bar{q}} \dot{\bar{q}} +
   \frac{\partial q(\bar{q},\dot{\bar{q}})}{\partial \dot{\bar{q}}} \ddot{\bar{q}} \right) .
\end{equation}
The idea behind a composite theory is that we might gain deeper understanding of a successful theory by recognizing more fundamental variables. We expect to reduce the number of physical degrees of freedom of the workhorse theory by introducing the more fundamental variables.

The passage from the Lagrangian formulation of composite higher derivative theories with linear dependence of $q$ on $\dot{\bar{q}}$ to Ostrogradsky's Hamiltonian formulation has been elaborated in \cite{hco235}. The main outcome of that work has been that, for this special class of higher derivative theories, there is a particularly simple and convenient way of formulating constraints. It has been verified explicitly that these constraints indeed eliminate all unstable solutions.

The goal of the present work is to develop an alternative to Ostrogradsky's general procedure that is more deeply adapted to the structure of composite theories. Rather than using $\bar{q}$ and $\dot{\bar{q}}$ as configurational variables and introducing the proper conjugate momentum variables in the Ostrogradsky approach, we here start from $\bar{q}$ and $q$ as the natural configurational variables and keep $p$ as the conjugate momentum of $q$. In this alternative procedure we keep the Hamiltonian structure of the workhorse theory and we can formulate the constraints even more naturally. We moreover recognize how solutions of a composite theory can be found by a two-step procedure.

The space $\bar{q}, q, \bar{p}, p$ is more naturally structured but larger than the space of the original Ostrogradsky approach (note that $q$ is a larger set of configurational variables than $\dot{\bar{q}}$). The new space has the appealing structure of the combined spaces of the original variables of the workhorse theory and the more fundamental variables used in the composition rule. The Hamiltonian formulation of higher derivative theories highlights that we are forced into larger spaces, where the combined space of the present formulation is particularly natural. Last, but definitely not least, the combined space provides the natural arena for the quantization of composite higher derivative theories.

\section{Basics of Lagrangian setting}
As a workhorse for our concrete development of a composite higher derivative theory, we consider a mechanical system with Lagrangian
\begin{equation}\label{Lagrangian}
   L(q,\dot{q}) = \frac{1}{2} m \dot{q}_i \dot{q}_i + \dot{q}_i \, u_i(q) - V(q) ,
\end{equation}
involving a discrete set of variables $q_i$, $i=1, \ldots I$, where $q$ stands for the list of all variables, $u_i(q)$, $V(q)$ are sufficiently smooth functions, $V(q)$ is bounded from below, and Einstein's summation convention is used (summation over indices occurring twice). For the composition rule we assume a linear dependence of $q$ on the time derivatives of the more fundamental variables $\bar{q}$,
\begin{equation}\label{variabletransf}
   q_i = \alpha_i(\bar{q}) + \beta_{ik}(\bar{q}) \, \dot{\bar{q}}_k ,
\end{equation}
where $\alpha_i(\bar{q})$, $\beta_{ik} (\bar{q})$ are sufficiently smooth functions of the variables $\bar{q}_k$, $k=1, \ldots K \leq I$ (more general dependencies on $\dot{\bar{q}}_k$ have been considered in \cite{DeriglazovKuznetsova07} with the goal of converting second class constraints in a Lagrangian setting). The variables $\dot{\bar{q}}_k$ for given $q$ and $\bar{q}$ can be extracted from Eq.~(\ref{variabletransf}) as
\begin{equation}\label{qbardotfct}
   \dot{\bar{q}}_k(q,\bar{q}) =
   \big[ q_i - \alpha_i(\bar{q}) \big] \, \beta_{il}(\bar{q}) B_{lk}(\bar{q}) ,
\end{equation}
where the square matrix $B_{lk}$ is the inverse of the symmetric $K \times K$ matrix $\beta_{ik} \beta_{il}$ (we assume that the $I \times K$  matrix $\beta_{ik}$ possesses the maximum possible rang, which is $K$). By inserting this result back into Eq.~(\ref{variabletransf}), we obtain the $I-K$ constraints
\begin{equation}\label{primaryconstraints}
   ( \delta_{ij} - {\cal P}_{ij} ) ( q_j - \alpha_j ) = 0 ,
\end{equation}
where
\begin{equation}\label{Projection}
   {\cal P}_{ij} = \beta_{ik} B_{kl} \beta_{jl}
\end{equation}
is a projection operator depending on $\bar{q}$. The constraints (\ref{primaryconstraints}) make sure that, for given $\bar{q}_k$, the $I$ variables $q_i$ do not contain more information than the $K$ variables $\dot{\bar{q}}_k$. In other words, these constraints turn $q_i$ and $\dot{\bar{q}}_k$ into equivalent sets of variables.

Equations (\ref{qbardotfct}) and (\ref{Projection}) can be written in a more compact form if we define the ``inverse'' matrix
\begin{equation}\label{betainverse}
   \beta^{-1}_{il} = \beta_{ik} B_{kl} ,
\end{equation}
which has the useful properties
\begin{equation}\label{betainverseprop}
   \beta^{-1}_{jk} \beta_{jl} = \delta_{kl} , \qquad
   \beta^{-1}_{il} \beta_{jl} = {\cal P}_{ij} .
\end{equation}
If we further introduce the derivatives
\begin{equation}\label{albeder}
   \alpha'_{ik}(\bar{q}) = \frac{\partial \alpha_i(\bar{q})}{\partial \bar{q}_k} , \qquad
   \beta'_{ilk}(\bar{q}) = \frac{\partial \beta_{il}(\bar{q})}{\partial \bar{q}_k} ,
\end{equation}
we obtain the identity
\begin{equation}\label{Projectionder}
   \frac{\partial {\cal P}_{ij}}{\partial \bar{q}_k} =
   \Big[ \beta^{-1}_{il} ( \delta_{j'j} - {\cal P}_{j'j} )
   + ( \delta_{ij'} - {\cal P}_{ij'} ) \beta^{-1}_{jl} \Big] \beta'_{j'lk} ,
\end{equation}
and the secondary constraints, which are associated with the preservation of the primary constraints (\ref{primaryconstraints}) in time, can be written as
\begin{equation}\label{secondaryconstraints}
   ( \delta_{ij} - {\cal P}_{ij} ) \big( \dot{q}_j - \alpha'_{jk} \dot{\bar{q}}_k
   - \beta'_{jkl} \, \dot{\bar{q}}_k \dot{\bar{q}}_l \big) = 0 .
\end{equation}

Stationarity of the action associated with the Lagrangian (\ref{Lagrangian}) with respect to variations of $q$ leads to the evolution equations of the workhorse theory,
\begin{equation}\label{Lagrangeqevol}
   m \ddot{q}_i + \omega_{ij} \dot{q}_j + \frac{\partial V}{\partial q_i} = 0 ,
\end{equation}
with
\begin{equation}\label{omegadef}
   \omega_{ij}(q) = \frac{\partial u_i(q)}{\partial q_j}
   - \frac{\partial u_j(q)}{\partial q_i}.
\end{equation}
If we restrict the variations of $q$ to those implied by variations of $\bar{q}$,
\begin{eqnarray}
   \delta I  &=& \delta q \left( \frac{\partial L}{\partial q}
   - \frac{d}{dt} \frac{\partial L}{\partial \dot{q}} \right) \nonumber \\
   &=& \delta \bar{q} \left( \frac{\partial q}{\partial \bar{q}}
   - \frac{d}{dt} \frac{\partial q}{\partial \dot{\bar{q}}} \right)
   \left( \frac{\partial L}{\partial q}
   - \frac{d}{dt} \frac{\partial L}{\partial \dot{q}} \right) ,
\label{actionvariations}
\end{eqnarray}
or if we equivalently require stationarity of the action associated with the Lagrangian (\ref{Lcomposite}), we obtain the following smaller set of evolution equations for the composite theory given by Eqs.~(\ref{Lagrangian}) and (\ref{variabletransf}),
\begin{equation}\label{Lagrangeqbarevol}
   \left ( \alpha'_{ik} + \beta'_{ilk} \, \dot{\bar{q}}_l
   - \frac{d}{dt} \beta_{ik} \right )
   \left ( m \ddot{q}_i + \omega_{ij} \dot{q}_j + \frac{\partial V}{\partial q_i} \right ) = 0 .
\end{equation}
Note that Eq.~(\ref{Lagrangeqbarevol}) contains third-order time derivatives of $q$ (the derivative operator $d/dt$ acts not only on $\beta_{ik}$, but also on all the subsequent expression in parentheses), implying a set of fourth-order differential equations for $\bar{q}$. This is the class of fourth-order differential equations considered in \cite{hco235}. We postpone the detailed discussion of constraints and their implications until we have developed the Hamiltonian setting.

\section{Passage to canonical Hamiltonian formulation}
We would now like to define Hamiltonian dynamics in a larger space based on the configurational variables $\bar{q}$ and $q$ (rather than $\bar{q}$ and $\dot{\bar{q}}$ in the Ostrogradsky approach). In addition to these configurational variables, we use the conjugate momenta of the workhorse theory,
\begin{equation}\label{conjmom}
   p_i = \frac{\partial L}{\partial \dot{q}_i} = m \dot{q}_i + u_i ,
\end{equation}
and the additional conjugate momenta $\bar{p}$ associated with $\bar{q}$, the physical meaning of which remains to be identified. Motivated by recovering Eq.~(\ref{qbardotfct}) as an evolution equation, we choose the Hamiltonian $H=H(\bar{q},q,\bar{p},p)$ as the following extension of the Hamiltonian for the workhorse theory,
\begin{equation}\label{Hamiltonian}
   H = \frac{[p_i - u_i(q)] [p_i - u_i(q)]}{2m} + V(q)
   + [ q_i - \alpha_i(\bar{q}) ] \, \beta^{-1}_{ik}(\bar{q}) \bar{p}_k .
\end{equation}
This Hamiltonian defines the composite theory in the enlarged space. It implies the following canonical evolution equations,
\begin{equation}\label{caneqmotqbar}
   \frac{d \bar{q}_k}{d t} = (q_i - \alpha_i) \, \beta^{-1}_{ik} ,
\end{equation}
\begin{equation}\label{caneqmotq}
   \frac{d q_i}{d t} = \frac{p_i - u_i}{m} ,
\end{equation}
\begin{equation}\label{caneqmotp}
   \frac{d}{d t} (p_i - u_i) = - \left( \omega_{ij} \, \frac{p_j - u_j}{m}
   + \frac{\partial V}{\partial q_i} + \beta^{-1}_{ik} \, \bar{p}_k  \right) ,
\end{equation}
and
\begin{eqnarray}
   \frac{d \bar{p}_k}{d t} &=& \big[ \alpha'_{ik} +
   (q_j - \alpha_j) \, \beta^{-1}_{jl} \, \beta'_{ilk} \big]
   \beta^{-1}_{in} \, \bar{p}_n \nonumber \\
   &-& (q_i - \alpha_i) ( \delta_{ij} - {\cal P}_{ij} ) \beta'_{jlk} B_{ln} \, \bar{p}_n  .
\label{caneqmotpbar}
\end{eqnarray}
Note that the term in the second line of Eq.~(\ref{caneqmotpbar}) vanishes if the primary constraints (\ref{primaryconstraints}) are imposed.

As intended by the choice of the Hamiltonian, the evolution equation (\ref{caneqmotqbar}) reproduces the relation for $\dot{\bar{q}}_k(q,\bar{q})$ given in Eq.~(\ref{qbardotfct}). Equation (\ref{caneqmotq}) coincides with the definition (\ref{conjmom}) of the conjugate momenta $p_i$. Equation (\ref{caneqmotp}) can be rewritten as
\begin{equation}\label{conjmombar}
   \beta^{-1}_{ik} \, \bar{p}_k =
   - \left ( m \ddot{q}_i + \omega_{ij} \dot{q}_j + \frac{\partial V}{\partial q_i} \right ) ,
\end{equation}
where Eq.~(\ref{caneqmotq}) has been used. Finally, by means of Eq.~(\ref{conjmombar}), the evolution equation (\ref{caneqmotpbar}) is found to be identical to Eq.~(\ref{Lagrangeqbarevol}) obtained from the Lagrangian approach. We have thus established a canonical Hamiltonian formulation of composite higher derivative theories on the enlarged space $(\bar{q},q,\bar{p},p)$, where the variables $\bar{q},q,p$ are a natural part of the setting for composite theories and $\bar{p}$ has been identified in Eq.~(\ref{conjmombar}).

We have already provided the primary and secondary constraints associated with the composition rule in Eqs.~(\ref{primaryconstraints}) and (\ref{secondaryconstraints}). Tertiary constraints result from the preservation of the secondary constraints in time, and so on, until eventually all constraints are preserved in time. The subsequent examples show that the constraints $\bar{p}_k = 0$ arise during this iteration procedure. These constraints keep the Hamiltonian (\ref{Hamiltonian}) bounded from below and hence eliminate the problem of instability. We hence suggest to elevate the constraints $\bar{p}_k = 0$, in addition to the constraints (\ref{primaryconstraints}), generally to the level of primary constraints which, according to Eq.~(\ref{caneqmotpbar}), do not lead to additional secondary constraints.

In short, the following scenario arises for composite higher derivative theories in the Hamiltonian approach. Dynamics takes place in a larger but canonically structured space, which is ideal for quantization. If the conjugate momenta $\bar{p}_k$ vanish initially, they vanish at all times. This constraint is recommended for guaranteeing stability \cite{hco235}. It implies that Eqs.~(\ref{caneqmotq}) and (\ref{caneqmotp}) are the evolution equations of the workhorse theory. The solution for $\bar{q}_k$ is then obtained from Eq.~(\ref{caneqmotqbar}) by post-processing. The constraints (\ref{primaryconstraints}) make sure that the resulting solution $\bar{q}_k$ is consistent with the composition rule (\ref{variabletransf}) of the composite theory.

It should be noted that the composition rule is the cornerstone of the proposed Hamiltonian approach. It provides the basic configurational variables $\bar{q}$ to supplement the variables $q$ of the workhorse theory and it leads directly to the primary constraints (\ref{primaryconstraints}), from which all further constraints arise by requiring dynamic invariance of all constraints.

\section{Example: harmonic oscillators} \label{secexample}
To gain a better understanding of the constraints and the elimination of instabilities, we consider two simple examples in more detail.

\subsection{Two oscillators} \label{secexample2osc}
As a first example, we consider two independent harmonic oscillators with mass $m$ and spring constants $h_1,h_2$ ($I=2$),
\begin{equation}\label{exLstandard2}
   L = \frac{1}{2} m \left( \dot{q}_1^2 + \dot{q}_2^2 \right)
   - \frac{1}{2} \left( h_1 q_1^2 + h_2 q_2^2 \right) .
\end{equation}
We express the coordinates $q_1,q_2$ in terms of $\bar{q}_1$ ($K=1$). For the functions in the composition rule (\ref{variabletransf}) we choose
\begin{equation}\label{alphabetaosc2}
   \alpha(\bar{q}) = \left(
              \begin{array}{c}
                \bar{q}_1 \\
                \bar{q}_1 \\
              \end{array}
            \right) , \qquad
   \beta = \lambda \left(
             \begin{array}{c}
               0 \\
               1 \\
             \end{array}
           \right) .
\end{equation}
For $\lambda \neq 0$, this choice for $\beta$ leads to
\begin{equation}\label{betainvPosc2}
   \beta^{-1} = \frac{1}{\lambda} \left(
             \begin{array}{c}
               0 \\
               1 \\
             \end{array}
           \right) , \qquad
   {\cal P} = \left(
                \begin{array}{cc}
                  0 & 0 \\
                  0 & 1 \\
                \end{array}
              \right) .
\end{equation}

Solution of the usual second-order equations (\ref{Lagrangeqevol}) for two harmonic oscillators [$i=1,2$; no summation over $i$ in Eqs.~(\ref{exLaguseq2}) and (\ref{solqi2})],
\begin{equation}\label{exLaguseq2}
   m \ddot{q}_i + h_i q_i = 0 ,
\end{equation}
requires four initial conditions. The solutions are of the form
\begin{equation}\label{solqi2}
   q_i = c_i \cos \omega_i t + c'_i \sin \omega_i t ,
\end{equation}
with the frequencies $\omega_i=\sqrt{h_i/m}$ and the four integration constants $c_i$, $c'_i$.

The system of Hamiltonian equations (\ref{caneqmotqbar})--(\ref{caneqmotpbar}) for the composite two-oscillator system become
\begin{equation}\label{Hamharmoscqbar2}
   \dot{\bar{q}}_1 = \frac{1}{\lambda} (q_2-\bar{q}_1) ,
\end{equation}
\begin{equation}\label{Hamharmoscq2}
   \dot{q}_1 = \frac{p_1}{m} , \quad
   \dot{q}_2 = \frac{p_2}{m} ,
\end{equation}
\begin{equation}\label{Hamharmoscp2}
   \dot{p}_1 = - h_1 q_1 , \quad
   \dot{p}_2 = - h_2 q_2 - \frac{1}{\lambda} \bar{p}_1 ,
\end{equation}
and
\begin{equation}\label{Hamharmoscpbar2}
   \dot{\bar{p}}_1 = \frac{1}{\lambda} \bar{p}_1 .
\end{equation}
This system consists of six first-order equations and hence requires six initial conditions. The complete solution of these linear equations in the large space can be given explicitly,
\begin{eqnarray}
   \bar{q}_1 &=& \bar{c}' \, e^{-t/\lambda}
   - \frac{\bar{c} \lambda}{2m} \frac{1}{1+\lambda^2\omega_2^2} \, e^{t/\lambda} \nonumber\\
   &+& \frac{(c_2 - \lambda\omega_2 c'_2) \cos \omega_2 t
   + (c'_2 + \lambda\omega_2 c_2) \sin \omega_2 t}{1+\lambda^2\omega_2^2} , \qquad
\label{sollargeqbar12}
\end{eqnarray}
\begin{equation}\label{sollargeq12}
   q_1 = c_1 \cos \omega_1 t + c'_1 \sin \omega_1 t ,
\end{equation}
\begin{equation}\label{sollargeq22}
   q_2 = c_2 \cos \omega_2 t + c'_2 \sin \omega_2 t
   - \frac{\bar{c} \lambda}{m} \frac{1}{1+\lambda^2\omega_2^2} \, e^{t/\lambda} ,
\end{equation}
\begin{equation}\label{sollargep12}
   p_1 = m \omega_1 ( c'_1 \cos \omega_1 t - c_1 \sin \omega_1 t ) ,
\end{equation}
\begin{equation}\label{sollargep22}
   p_2 = m \omega_2 ( c'_2 \cos \omega_2 t - c_2 \sin \omega_2 t )
   - \frac{\bar{c}}{1+\lambda^2\omega_2^2} \, e^{t/\lambda} ,
\end{equation}
and
\begin{equation}\label{sollargepbar12}
   \bar{p}_1 = \bar{c} \, e^{t/\lambda} .
\end{equation}
Note the occurrence of exponentially increasing (and decreasing) terms in these solutions.

The linear equations (\ref{Hamharmoscqbar2})--(\ref{Hamharmoscpbar2}) should actually be solved subject to the primary constraint (\ref{primaryconstraints}), which becomes
\begin{equation}\label{constraintsosc12}
   q_1 = \bar{q}_1 .
\end{equation}
Consistency of this constraint with the time evolution leads to three additional constraints,
\begin{equation}\label{constraintsosc22}
  p_1 = \frac{m}{\lambda} (q_2 - q_1) , \quad
  p_2 = p_1 - \lambda h_1 q_1
\end{equation}
and, moreover,
\begin{equation}\label{constraintsosc32}
   \bar{p}_1 = \lambda (h_1-h_2) q_2 .
\end{equation}
For $h_1 \neq h_2$, further constraints enforce the trivial solution $\bar{q}_1=q_1=q_2=p_1=p_2=\bar{p}_1=0$. For $h_1=h_2$, the four constraints (\ref{constraintsosc12})--(\ref{constraintsosc32}) reduce the number of free parameters in the solutions from six to two. In this case, the two-parameter solution $q_1=\bar{q}_1$ is given in Eq.~(\ref{solqi2}) or (\ref{sollargeq12}), and $q_2$ is obtained as $q_2=\bar{q}_1+\lambda\dot{\bar{q}}_1$. In either case, the hierarchy of constraints ends with $\bar{p}_1=0$, which is the crucial constraint for eliminating instabilities.

The same conclusions can be reached by considering the explicit solutions (\ref{sollargeqbar12})--(\ref{sollargepbar12}). The composition rule $q_1 = \bar{q}_1$ implies that nonvanishing solutions can be obtained only for $\bar{c}=\bar{c}'=0$ (suppressing all exponentially growing or decreasing terms) and $\omega_1=\omega_2$ (implying $h_1=h_2$). The matching conditions
$c_1 (1+\lambda^2\omega_2^2) = c_2 - \lambda\omega_2 c'_2$
and
$c'_1 (1+\lambda^2\omega_2^2) = c'_2 + \lambda\omega_2 c_2$
reduce the number of free initial values to two.

\subsection{Three oscillators}
We next revisit the example of three harmonic oscillators discussed in detail in Appendix~A of \cite{hco235}. Starting from three independent harmonic oscillators with mass $m$ and spring constant $h$ ($I=3$),
\begin{equation}\label{exLstandard}
   L = \frac{1}{2} m \left( \dot{q}_1^2 + \dot{q}_2^2 + \dot{q}_3^2 \right)
   - \frac{1}{2} h \left( q_1^2 + q_2^2 + q_3^2 \right) ,
\end{equation}
we express the coordinates $q_1,q_2,q_3$ in terms of $\bar{q}_1,\bar{q}_2$ ($K=2$). For the functions in the composition rule (\ref{variabletransf}) we choose
\begin{equation}\label{alphabetaosc}
   \alpha(\bar{q}) = \left(
              \begin{array}{c}
                \bar{q}_1 \\
                \bar{q}_2 \\
                0 \\
              \end{array}
            \right) , \qquad
   \beta = \lambda \left(
             \begin{array}{cc}
               0 & 1 \\
               0 & 0 \\
               1 & 0 \\
             \end{array}
           \right) .
\end{equation}
For $\lambda \neq 0$, the composition rule is a one-to-one transformation between the variables $q_i$, $\dot{q}_i$ and $\bar{q}_k$, $\dot{\bar{q}}_k$, $\ddot{\bar{q}}_k$. We moreover find
\begin{equation}\label{betainvPosc}
   \beta^{-1} = \frac{1}{\lambda} \left(
             \begin{array}{cc}
               0 & 1 \\
               0 & 0 \\
               1 & 0 \\
             \end{array}
           \right) , \qquad
   {\cal P} = \left(
                \begin{array}{ccc}
                  1 & 0 & 0 \\
                  0 & 0 & 0 \\
                  0 & 0 & 1 \\
                \end{array}
              \right) .
\end{equation}

Solution of the usual second-order equations (\ref{Lagrangeqevol}) for three harmonic oscillators,
\begin{equation}\label{exLaguseq}
   m \ddot{q}_i + h q_i = 0 ,
\end{equation}
requires six initial conditions. The solutions are of the form
\begin{equation}\label{solqi}
   q_i = c_i \cos \omega t + c'_i \sin \omega t ,
\end{equation}
with the frequency $\omega=\sqrt{h/m}$ and the six integration constants $c_i$, $c'_i$.

The system of Hamiltonian equations (\ref{caneqmotqbar})--(\ref{caneqmotpbar}) for the composite theory,
\begin{equation}\label{Hamharmoscqbar}
   \dot{\bar{q}}_1 = \frac{1}{\lambda} q_3 , \quad
   \dot{\bar{q}}_2 = \frac{1}{\lambda} (q_1-\bar{q}_1) ,
\end{equation}
\begin{equation}\label{Hamharmoscq}
   \dot{q}_1 = \frac{p_1}{m} , \quad
   \dot{q}_2 = \frac{p_2}{m} , \quad
   \dot{q}_3 = \frac{p_3}{m} ,
\end{equation}
\begin{equation}\label{Hamharmoscp}
   \dot{p}_1 = - h q_1 - \frac{\bar{p}_2}{\lambda} , \quad
   \dot{p}_2 = - h q_2 , \quad
   \dot{p}_3 = - h q_3 - \frac{\bar{p}_1}{\lambda} ,
\end{equation}
and
\begin{equation}\label{Hamharmoscpbar}
   \dot{\bar{p}}_1 = \frac{1}{\lambda} \bar{p}_2 , \quad
   \dot{\bar{p}}_2 = 0 ,
\end{equation}
consists of ten first-order equations and hence requires ten initial conditions. The explicit solution in the large space has been given in Appendix~A of \cite{hco235}. We here restrict ourselves to solving these linear equations subject to the primary constraint (\ref{primaryconstraints}), which becomes
\begin{equation}\label{constraintsosc1}
   q_2 = \bar{q}_2 .
\end{equation}
Consistency of this constraint with the time evolution leads to five additional constraints,
\begin{equation}\label{constraintsosc2}
  p_2 = \frac{m}{\lambda} (q_1 - \bar{q}_1) , \quad
  p_1 = \frac{m}{\lambda} q_3 - \lambda h \bar{q}_2 , \quad
  p_3 = - \lambda h \bar{q}_1
\end{equation}
and ultimately
\begin{equation}\label{constraintsosc3}
   \bar{p}_1 = 0 , \quad \bar{p}_2 = 0 .
\end{equation}
The six constraints (\ref{constraintsosc1})--(\ref{constraintsosc3}) reduce the number of parameters from ten to four. The constraints (\ref{constraintsosc3}) imply that Eqs.~(\ref{exLaguseq}) and (\ref{Hamharmoscp}) coincide, which is the generally anticipated consequence of $\bar{p}_k=0$. From the solutions (\ref{solqi}), we obtain
\begin{equation}\label{solqbar}
   \bar{q}_1 = q_1 - \lambda \dot{q}_2 , \quad  \bar{q}_2 = q_2 ,
\end{equation}
which indeed contain only four parameters. The remaining constraint
$q_3 = \lambda \dot{\bar{q}}_1 = \lambda \dot{q}_1 - \lambda^2 \ddot{q}_2$
shows that the solution for $q_3$ does not contain independent initial conditions. The number of independent harmonic oscillators is reduced from three to two.

Finally, the constraints (\ref{constraintsosc3}) imply that the Hamiltonian (\ref{Hamiltonian}) is bounded from below. These constraints suppress all exponentially increasing or decreasing terms in the solutions (cf.\ Sect.~\ref{secexample2osc}, or see Appendix~A of \cite{hco235} for details), so that we have no problems with instabilities for the enlarged system.

\section{Conclusions and outlook}
We have recast composite higher derivative theories, which are conveniently introduced in a Lagrangian setting through a workhorse theory and a composition rule for the variables, into a canonical Hamiltonian form. The canonical variables $q$, $p$ of the workhorse theory are supplemented by the more basic variables $\bar{q}$ of the composition rule, and the corresponding conjugate momenta $\bar{p}$ vanish if the equations of motion of the workhorse theory are satisfied. Dynamic invariance of the primary constraints, which express the consistency of the composition rule, eventually leads to the constraints $\bar{p}=0$. These constraints eliminate instabilities and imply that the solutions of the composite theory are restricted solutions of the workhorse theory. These restricted solutions can be determined in a post-processing step.

The Lagrangian and Hamiltonian formulations of composite theories for mechanical systems can easily be generalized to field theories. The motivation for the present work has actually been the development of a higher derivative theory of gravity by employing the Yang-Mills theory based on the Lorentz group (see, e.g., Chap.~15 of \cite{WeinbergQFT2} or Sect.~15.2 of \cite{PeskinSchroeder}) as a workhorse. The composition rule is a transformation from the tetrad decomposition of the metric of general relativity to the vector potential of the Yang-Mills theory. For the linearized version of the resulting theory, gravitational waves have been discussed in Appendix~B of \cite{hco235}.

The canonical Hamiltonian formulation of composite theories is ideal for the purpose of quantization. As for gauge theories, an elegant formulation arises in an enlarged space so that constraints have to be introduced in the quantization procedure. BRST quantization \cite{Nemeschanskyetal86,PeskinSchroeder,hco229}, where the acronym BRST refers to Becchi, Rouet, Stora \cite{BecchiRouetStora76} and Tyutin \cite{Tyutin75}, is a well-established and powerful tool for quantization under constraints. Dissipative quantum field theory \cite{hco200,hcoqft} provides an ideal setting for the BRST quantization of a Yang-Mills based composite theory of gravity.

In this work, composite theories have been formulated in the setting or reversible equations. They may be considered as the passage from a more detailed to a less detailed level of description. The Hamiltonian formulation reveals that this passage is achieved most elegantly by considering both levels simultaneously and then narrowing down the combined space by constraints.

In general, the passage from more detailed to less detailed levels of description requires an entropy accounting for the eliminated degrees of freedom and leads to the emergence of irreversibility \cite{hcobet}. Also for irreversible systems, the consideration of the combined space is a powerful tool for elaborating the relation between two levels of description \cite{Grmela10a,Grmela14,Grmela17,Grmela18,Grmelaetal19}. A detailed comparison of the ideas for irreversible systems with the construction of composite reversible theories proposed in the present work might be illuminating.


\end{document}